\documentclass[printer]{aa}
\usepackage{graphicx,natbib}
\usepackage{txfonts}
\usepackage{epic,eepic}

\usepackage{color}
\definecolor{red}{rgb}{0.,0.,0.0}

\begin{document}
 
%
%
\def\valid{}    

\font\caps=cmcsc10                  
\font\dunh=cmdunh10  at 12.0 true pt 
\font\dunhs=cmdunh10 
\font\vbold=cmbx10 scaled \magstep1 
\font\sevenbf=cmbx7
\font\sevenit=cmti7
\font\Kapi=cmr17

\def\MEV{DOME}
\def\RTE{equation of radiative transfer}
\def\etal{{et al}}
\def\HW{H\&W}
\def\OK{O\&K}
\def\ok{O\&K}
\def\RH{R\&H}

\def\ibmrs{\hbox{\tt RS/6000}}
\def\hp{\hbox{\tt HP~9000}}
\def\dec{\hbox{\tt DEC~5000}}
\def\axp{\hbox{\tt AXP}}
\def\ibmmf{\hbox{\tt IBM~3090}}
\def\ibmpc{\hbox{\tt 486DX}}
\def\cray{\hbox{\tt Cray 2}}
\def\ymp{\hbox{\tt YMP}}
\def\nec{\hbox{\tt NEC}}

\def\g{\gamma}
\def\b{\beta}
\def\m{\mu}
\def\e{\epsilon}
\def\n{\nu}
\def\l{\lambda}
\def\L{\Lambda}
\def\t{\tau}
\def\pder#1#2{{\partial #1 \over \partial #2}}
\def\div#1#2{{#1\over #2}}
\def\rout{\ifmmode{r_{\rm out}}\else\hbox{$r_{\rm out}$}\fi}
\def\tmax{\ifmmode{\tau_{\rm max}}\else\hbox{$\tau_{\rm max}$}\fi}
\def\tstd{\ifmmode{\tau_{\rm std}}\else\hbox{$\tau_{\rm std}$}\fi}
\def\vmax{\ifmmode{v_{\rm max}}\else\hbox{$v_{\rm max}$}\fi}
\def\muE{\ifmmode{\mu_{\rm E}}\else\hbox{$\mu_{\rm E}$}\fi} 
\def\pE{\ifmmode{p_{\rm E}}\else\hbox{$p_{\rm E}$}\fi} 
\def\bmax{\ifmmode{\b_{\rm max}}\else\hbox{$\b_{\rm max}$}\fi}
\def\kms{\hbox{$\,$km$\,$s$^{-1}$}}
\def\ergs{\hbox{$\,$erg$\,$s$^{-1}$}}
\def\kpc{\hbox{$\,$kpc} }
\def\ang{\hbox{\AA}}
\def\Msun{\hbox{$\,$M$_\odot$} }
\def\Lsun{\hbox{$\,$L$_\odot$} }
\def\Teff{\hbox{$\,T_{\rm eff}$} }
\def\alog#1{\times 10^{#1}}
\def\rin{\hbox{$r_{\rm in}$} }
\def\rout{\hbox{$r_{\rm out}$} }

\def\lstar{\ifmmode{\Lambda^*}\else\hbox{$\Lambda^*$}\fi} 
\def\Lstar{\ifmmode{\Lambda^*}\else\hbox{$\Lambda^*$}\fi} 
\def\Rop{\ifmmode{[R_{ij}]}\else\hbox{$[R_{ij}]$}\fi}
\def\Rij{\Rop}
\def\Rji{\ifmmode{[R_{ji}]}\else\hbox{$[R_{ji}]$}\fi}
\def\Rstar{\ifmmode{[R_{ij}^*]}\else\hbox{$[R_{ij}^*]$}\fi}
\def\Rijstar{\Rstar}
\def\Rjistar{\ifmmode{[R_{ji}^*]}\else\hbox{$[R_{ji}^*]$}\fi}
\def\DRji{\ifmmode{[\Delta R_{ji}]}\else\hbox{$[\Delta R_{ji}]$}\fi}
\def\DRij{\ifmmode{[\Delta R_{ij}]}\else\hbox{$[\Delta R_{ij}]$}\fi}

\def\Jb{{\bar J}}
\def\Jbar{{\bar J}}
\def\Jnew{{\bar J_{\rm new}}}
\def\Jold{{\bar J_{\rm old}}}
\def\Jfs{{\bar J_{\rm fs}}}
\def\Snew{{S_{\rm new}}}
\def\Sold{{S_{\rm old}}}
\def\Amat{\mat{A}}             

\def\ns{\ifmmode{N_{\rm s}}          
        \else\hbox{$N_{\rm s}$}\fi}
\def\ion#1{\hbox{ #1}}         

\def\peq{\mathbin{\hbox{$+$}\hbox{$=$}}}

\def\mat#1{{\bf #1}}     
\def\vek#1{{#1}}         

\newcount\eqcount
\eqcount=0
\def
  \nummer{
    \global\advance\eqcount by 1
    (\the\eqcount)
  }

\def
  \numadv{
    \global\advance\eqcount by 1
  }

\def
   \numout#1{
     (\the\eqcount #1)
  }

\def\ivek#1#2{\ifmmode{\vek{I}^{#1}_{#2}}
        \else\hbox{$\vek{I}^{#1}_{#2}$}\fi}

\def\ip#1{\ivek{+}{#1}}      
\def\im#1{\ivek{-}{#1}}      

\def\tmat#1#2{\ifmmode{\mat{t}^{#1}_{#2}}
        \else\hbox{$\mat{t}^{#1}_{#2}$}\fi}
\def\rmat#1#2{\ifmmode{\mat{r}^{#1}_{#2}}
        \else\hbox{$\mat{r}^{#1}_{#2}$}\fi}
\def\bvek#1#2{\ifmmode{\beta^{#1}_{#2}}
        \else\hbox{$\beta^{#1}_{#2}$}\fi}

\def\tpi#1{\tmat{+}{#1}}
\def\tmi#1{\tmat{-}{#1}}
\def\rmi#1{\rmat{-}{#1}}
\def\rpi#1{\rmat{+}{#1}}
\def\bpi#1{\bvek{+}{#1}}
\def\bmi#1{\bvek{-}{#1}}

\def\tp{\tmat{+}{}}          
\def\tm{\tmat{-}{}}          
\def\rmm{\rmat{-}{}}         
\def\rp{\rmat{+}{}}          
\def\bp{\bvek{+}{}}          
\def\bm{\bvek{-}{}}          
\def\tpm{\tmat{\pm}{}}       
\def\rpm{\rmat{\pm}{}}       
\def\bpm{\bvek{\pm}{}}       

\def\lp{\ifmmode{\lambda^+_\tau}           
        \else\hbox{$\lambda^+_\tau$}\fi}
\def\lm{\ifmmode\lambda^-_\tau             
        \else\hbox{$\lambda^-_\tau$}\fi}

%
%
%
%



\def\aasref@jnl#1{{\rm #1}}

\def\aj{\aasref@jnl{AJ}}                   
\def\araa{\aasref@jnl{ARA\&A}}             
\def\apj{\aasref@jnl{ApJ}}                 
\def\apjl{\aasref@jnl{ApJ}}                
\def\apjs{\aasref@jnl{ApJS}}               
\def\ao{\aasref@jnl{Appl.~Opt.}}           
\def\apss{\aasref@jnl{Ap\&SS}}             
\def\aap{\aasref@jnl{A\&A}}                
\def\aapr{\aasref@jnl{A\&A~Rev.}}          
\def\aaps{\aasref@jnl{A\&AS}}              
\def\azh{\aasref@jnl{AZh}}                 
\def\baas{\aasref@jnl{BAAS}}               
\def\jrasc{\aasref@jnl{JRASC}}             
\def\memras{\aasref@jnl{MmRAS}}            
\def\mnras{\aasref@jnl{MNRAS}}             
\def\pra{\aasref@jnl{Phys.~Rev.~A}}        
\def\prb{\aasref@jnl{Phys.~Rev.~B}}        
\def\prc{\aasref@jnl{Phys.~Rev.~C}}        
\def\prd{\aasref@jnl{Phys.~Rev.~D}}        
\def\pre{\aasref@jnl{Phys.~Rev.~E}}        
\def\prl{\aasref@jnl{Phys.~Rev.~Lett.}}    
\def\pasp{\aasref@jnl{PASP}}               
\def\pasj{\aasref@jnl{PASJ}}               
\def\qjras{\aasref@jnl{QJRAS}}             
\def\skytel{\aasref@jnl{S\&T}}             
\def\solphys{\aasref@jnl{Sol.~Phys.}}      
\def\sovast{\aasref@jnl{Soviet~Ast.}}      
\def\ssr{\aasref@jnl{Space~Sci.~Rev.}}     
\def\zap{\aasref@jnl{ZAp}}                 
\def\nat{\aasref@jnl{Nature}}              
\def\iaucirc{\aasref@jnl{IAU~Circ.}}       
\def\aplett{\aasref@jnl{Astrophys.~Lett.}} 
\def\apspr{\aasref@jnl{Astrophys.~Space~Phys.~Res.}}
\def\bain{\aasref@jnl{Bull.~Astron.~Inst.~Netherlands}} 
\def\fcp{\aasref@jnl{Fund.~Cosmic~Phys.}}  
\def\gca{\aasref@jnl{Geochim.~Cosmochim.~Acta}}   
\def\grl{\aasref@jnl{Geophys.~Res.~Lett.}} 
\def\jcp{\aasref@jnl{J.~Chem.~Phys.}}      
\def\jgr{\aasref@jnl{J.~Geophys.~Res.}}    
\def\jqsrt{\aasref@jnl{J.~Quant.~Spec.~Radiat.~Transf.}}
\def\memsai{\aasref@jnl{Mem.~Soc.~Astron.~Italiana}}
\def\nphysa{\aasref@jnl{Nucl.~Phys.~A}}   
\def\physrep{\aasref@jnl{Phys.~Rep.}}   
\def\physscr{\aasref@jnl{Phys.~Scr}}   
\def\planss{\aasref@jnl{Planet.~Space~Sci.}}   
\def\procspie{\aasref@jnl{Proc.~SPIE}}   

\let\astap=\aap
\let\apjlett=\apjl
\let\apjsupp=\apjs
\let\applopt=\ao

\baselineskip=12pt

\title{A 3D radiative transfer framework: IV. spherical \& cylindrical coordinate systems}

\titlerunning{3D radiative transfer framework IV}
\authorrunning{Hauschildt and Baron}
\author{Peter H. Hauschildt\inst{1} and E.~Baron\inst{1,2,3}}

\institute{
Hamburger Sternwarte, Gojenbergsweg 112, 21029 Hamburg, Germany;
yeti@hs.uni-hamburg.de 
\and
Dept. of Physics and Astronomy, University of
Oklahoma, 440 W.  Brooks, Rm 100, Norman, OK 73019 USA;
baron@ou.edu
\and
Computational Research Division, Lawrence Berkeley National Laboratory, MS
50F-1650, 1 Cyclotron Rd, Berkeley, CA 94720-8139 USA
}

\date{Received date \ Accepted date}

\abstract
{}
{We extend our framework for 3D radiative transfer calculations
with a non-local operator splitting methods along (full) characteristics
to spherical and cylindrical coordinate systems. These coordinate systems
are better suited to a number of physical problems than Cartesian 
coordinates.
}
{The scattering problem for line transfer is solved via means of an
  operator splitting (OS) technique. The formal solution
  is based on a full characteristics method. The approximate
  $\Lambda$ operator is constructed considering nearest
  neighbors exactly. The code is parallelized over both wavelength and solid angle
  using the MPI library.
}
{We present the results of several test cases with different values of
  the thermalization parameter for the different coordinate systems.
 The results are directly compared to 1D plane parallel
  tests. The 3D results agree very
  well with the well-tested 1D calculations.
}
{Advances in modern computers will make realistic 3D radiative transfer
  calculations possible in the near future. 
}

\keywords{Radiative transfer -- Scattering}

\maketitle

\section{Introduction}

In 
\citet[][hereafter: Paper I]{3drt_paper1},
\citet[][hereafter: Paper II]{3drt_paper2} and
\cite[][hereafter: Paper III]{3drt_paper3}  we described a framework for
the solution of the radiative transfer equation for scattering continua
and lines
in 3D (when we say 3D we mean three spatial dimensions, plus three
momentum dimensions) for 
the time independent, static case using full characteristics
based on voxels in Cartesian coordinates. Cartesian coordinates
are by no means an optimal choice for many 3D physical systems,
for example supernova atmospheres are better described in 3D
spherical and accretion disks are better described in 3D
cylindrical coordinates. These coordinate systems make it 
a little harder and computationally more expensive to track the characteristics through the 
curvilinear voxel grid, however, this is more than overcompensated
by the higher efficiency of voxel usage, for example consider
putting a roughly spherical structure where density and
composition vary strongly with radius into a Cartesian coordinate
system cube.

{We describe our method, its rate of convergence, and present
  comparisons to our 
  well-tested 1-D calculations.}

\section{Method}

In the following discussion we use  notation of Papers I -- III.  The
basic framework and the methods used for the formal solution and the solution
of the scattering problem via operator splitting are discussed in detail in
Papers I and II and will not be repeated here. 

We use a ``full characteristics'' approach that {completely} tracks a set of
characteristics of the radiative transfer equation {from the inner or outer
boundary} through the computational domain {to their exit voxels and takes care
that each voxel is hit by at least one characteristic per solid angle}.  One
characteristic is started at each boundary voxel and then tracked through the
voxel grid until it leaves the other boundary.  The direction of a bundle of
characteristics is determined by a set of solid angles $(\theta,\varphi)$ which
correspond to a normalized momentum space vector $(p_x,p_y,p_z)$.

In contrast to the case of Cartesian coordinates the volumes of the voxels can
be vastly different in spherical and Cartesian coordinate systems. Therefore,
tracking a characteristic across the voxel grid is done using an adaptive
stepsize control along the characteristic so that a characteristic steps from
one voxel to its neighbor in the direction of the characteristic. In addition,
we make sure that each voxel is hit by at least one characteristic for 
each solid angle, which is important for the inner (small radii) voxels.

The code is parallelized as described in Papers I and II.

\section{spherical coordinates}

\subsection{Testing environment}
We use the framework discussed in Paper I and II as the baseline for the
line transfer problems discussed in this paper. 
Our basic setup is similar to that discussed in Paper II.  
We use a
sphere with a grey continuum opacity parameterized by a power 
law in the continuum optical depth $\tstd$. The basic model parameters
are
\begin{enumerate}
\item Inner radius $r_c=10^{10}\,$cm, outer radius $\rout = 2\alog{10}\,$cm (large test
case) or  $\rout = 1.1\alog{10}\,$cm (small test case).
\item Minimum optical depth in the continuum $\t_{\rm std}^{\rm min} =
10^{-4}$ and maximum optical depth in the continuum $\t_{\rm std}^{\rm
max} = 10^{8}$.
\item Constant temperature with $\Teff=10^4$~K, therefore, the Planck function $B$ is also constant.
\item Outer boundary condition $I_{\rm bc}^{-} \equiv 0$ and diffusion
inner boundary condition for all wavelengths.
\item Continuum extinction $\chi_c = C/r^6$, with the constant $C$
fixed by the radius and optical depth grids, for the continuum tests,
$\chi_c = C$ for the line transfer tests
\item Parameterized coherent \& isotropic continuum scattering by
defining
\[
\chi_c = \epsilon_c \kappa_c + (1-\epsilon_c) \sigma_c
\]
with $0\le \epsilon_c \le 1$. 
$\kappa_c$ and $\sigma_c$ are the 
continuum absorption and scattering coefficients.
\end{enumerate}
This problem is used because the results can be directly compared
with the results obtained with a 1D spherical radiation transport
code \cite[]{s3pap} to assess the accuracy of the method. 
The sphere is put at the center of the 3D grid. The
radial grid in the 3D coordinates is logarithmic in $\tstd$ and
has, therefore, highly variable spacing in $r$. In 3D spherical
coordinates $(r,\theta_c,\phi_c)$ 
(note: the subscript $c$ indicates
a coordinate in order to distinguish with the solid angle
$\Omega = (\theta,\phi)$) we use regular grid in $(\theta_c,\phi_c)$ with 
resolutions given below. In the tests with cylindrical
coordinates $(\rho,\phi_c,z)$ we use a regular grid in $z$ and $\phi_c$
and a variable grid in $\rho$.
The solid angle space was discretized in $(\theta,\phi)$ with $n_\theta=n_\phi$
if not stated otherwise. In the following we discuss the results of various
tests. In all tests we use the full characteristics method for the 3D RT
solution.

The line of the simple 2-level model atom is parameterized by the ratio of the
profile averaged  line opacity $\chi_l$ to the continuum opacity $\chi_c$ and
the line thermalization parameter $\epsilon_l$. For the test cases presented
below, we have used $\chi_c = {\rm const}$, $\epsilon_c=1$, and a grey temperature structure.
We set $\chi_l/\chi_c = 10^6$ to simulate a strong line, with
varying $\epsilon_l$ (see below). With this setup, the optical depths as seen
in the line range from $10^{-2}$ to $10^6$. We use 32 wavelength points to model
the full line profile, including wavelengths outside the line for the
continuum.  We did not require the line to thermalize at the center of the test
configurations, this is a typical situation for a full 3D
configuration as the location (or even existence) of the
thermalization depths becomes more 
ambiguous than in the 1D case.

In the following we discuss the results of various tests.  Unless otherwise
stated, the tests were run on parallel computers using 4 to 512 CPUs (depending
on the test). In spherical coordinates we use a voxel grid with $(n_r,n_{\theta_c},
n_{\phi_c}) = (65,33,65)$ voxels, this corresponds to an angular resolution of
about 5.5 degrees in $\theta_c$ and $\phi_c$. Projected on the surface of the Earth,
this means that Europe is covered by about 43 voxels. The spherical test cases
use incoming intensities of zero at the outside ($R_{\rm max})$ and $I=B$ at 
the inside (($R_{\rm min})$.

\subsection{Results}

\subsubsection{Continuum tests}

\begin{figure}
\centering
\resizebox{\hsize}{!}{\includegraphics[angle=90]{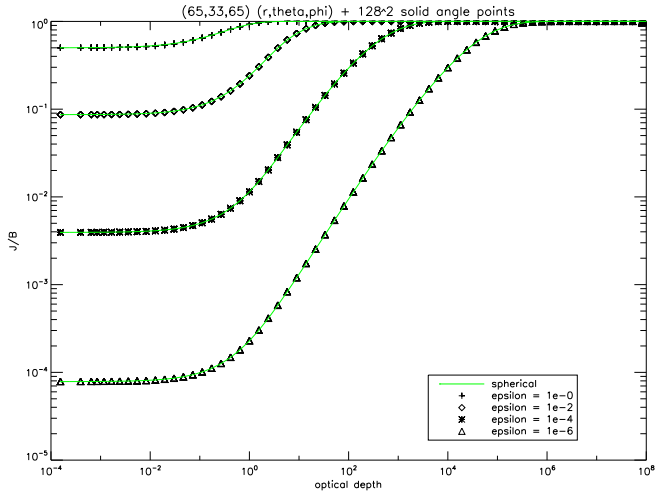}}
\caption{\label{fig:cont} The ratio of the mean intensity $J$ and the 
Planck function $B$ as function of the radial optical depth $\tstd$ 
for the 'large' test case in spherical coordinates and different values $\epsilon_c$. The optical depth is measured along the 
radius $r$. 
}
\end{figure}

In Fig.~\ref{fig:cont} we show the results for 3D continuum transfer 
with 3D spherical coordinates for various values of the thermalization
parameter $\epsilon_c$. The results are compared to 1D results (full line).
In all cases, the agreement is excellent, even for the case with $\epsilon_c=10^{-6}$.
Compared to paper I the agreement is better, this is caused by the better
discretization of a spherical configuration in spherical coordinates,
rather than fitting a sphere into a Cartesian grid. It illustrates the 
importance of adapted coordinate systems to obtain the optimal mix
between computational efficiency and numerical accuracy.

\begin{figure}
\centering
\resizebox{\hsize}{!}{\includegraphics[angle=00]{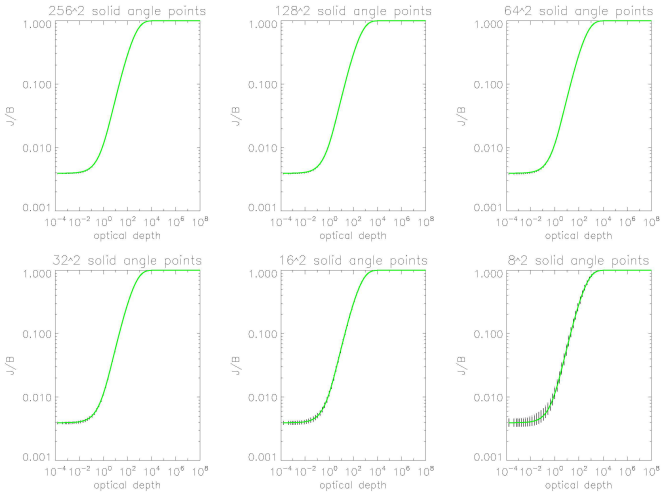}}
\caption{\label{fig:cont:omega:small} Effect of solid angle 
resolution on the results of the 3D calculations. The plots
show the results for the 'small' continuum test case in spherical coordinates with $\epsilon_c=10^{-4}$ 
and for various solid angle resolutions with $\n_\theta = n_\varphi$ from
$(256,256)$ to $(8,8)$. Full line: results of the 1D calculation, dots: 
results from the 3D calculations.
}
\end{figure}

\begin{figure}
\centering
\resizebox{\hsize}{!}{\includegraphics[angle=00]{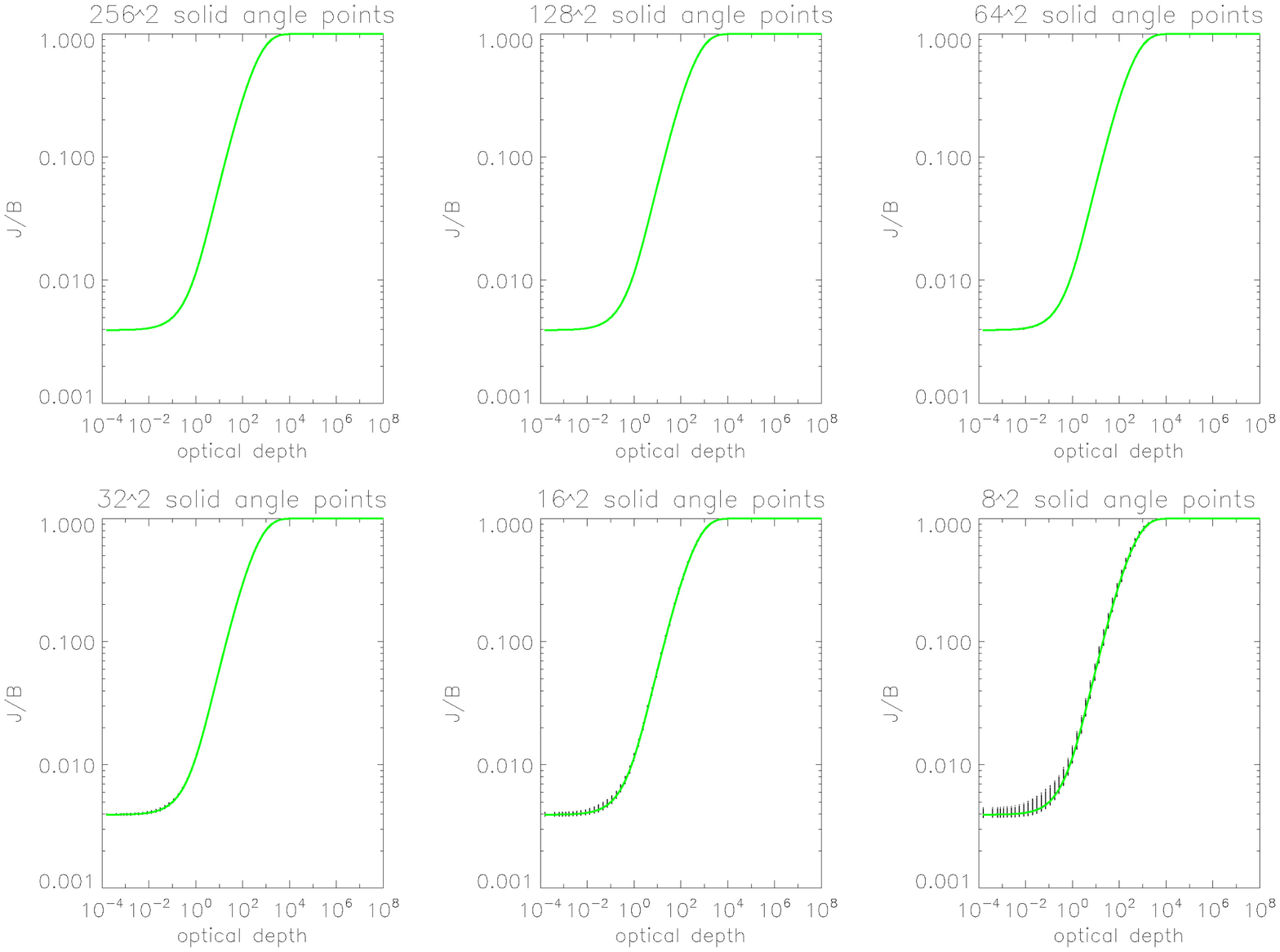}}
\caption{\label{fig:cont:omega:large} Effect of solid angle 
resolution on the results of the 3D calculations in spherical coordinates. The plots
show the results for the 'large' continuum test case with $\epsilon_c=10^{-4}$ 
and for various solid angle resolutions with $\n_\theta = n_\varphi$ from
$(256,256)$ to $(8,8)$. Full line: results of the 1D calculation, dots: 
results from the 3D calculations.
}
\end{figure}

Good coverage of the solid angle is important for 3D transfer. The 1D
code automatically adjusts the angle coverage related to the geometry of the
configuration, however, in the 3D code the solid angles have to be prescribed.
In Fig.~\ref{fig:cont:omega:small} we demonstrate the importance of solid angle
sampling for a 'small' test case with $\epsilon_c=10^{-4}$. The plots show 
the ratio $J/B$ as function of radial optical depth for various solid 
angle samplings and for all $(\phi_c,\theta_c)$ compared to the 1D results. 
It is immediately clear that for the 'small' test case a solid angle sampling of 
$64^2$ points or more is  reasonable, but that $16^2$ or less points result
in a considerable scatter of the results. This result does not depend
strongly on the radial extension of the configuration, as shown in 
Fig.~\ref{fig:cont:omega:large}.

\begin{figure}
\centering
\resizebox{\hsize}{!}{\includegraphics[angle=90]{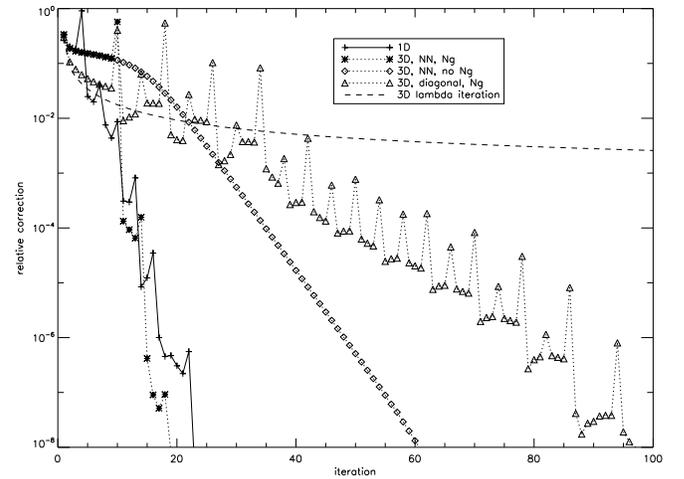}}
\caption{\label{fig:cont:convergence} Convergence behavior for the 'large' continuum
test case in spherical coordinates with $\epsilon_c=10^{-4}$ for various combinations of local/non-local $\Lstar$
and Ng acceleration. 'NN' indicates the non-local nearest neighbor $\Lstar$ operator,
'diagonal' the local $\Lstar$. For comparison the results for the 1D spherical
code and the 3D lambda iterations are also plotted.
}
\end{figure}

\begin{figure}
\centering
\resizebox{\hsize}{!}{\includegraphics[angle=90]{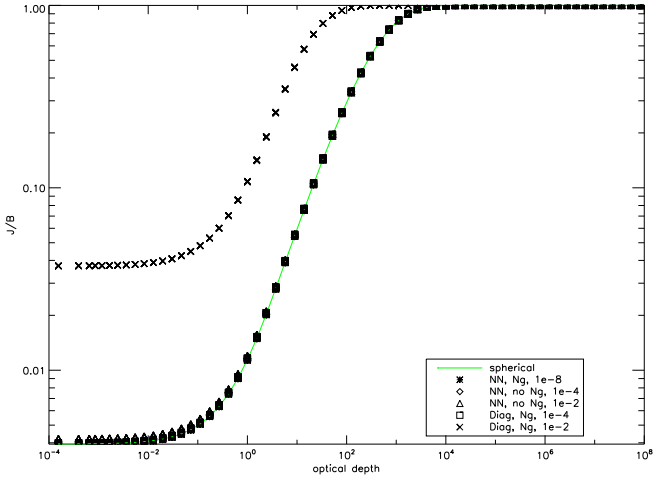}}
\caption{\label{fig:cont:relax} Effects of relaxing the convergence criterions
on the results of the 3D calculations in spherical coordinates. The plots show the results for the
'large' continuum test case with $\epsilon_c=10^{-4}$ and for various convergence
limits for nearest neighbor (NN) and diagonal (Diag) $\Lstar$ operators with
and without Ng acceleration. The numbers are the maximum allowed relative
change over all voxels used to halt the iterations.  Full line: results of the
1D calculation, dots: results from the 3D calculations.
}
\end{figure}

The convergence rate of the nearest neighbor $\Lstar$ ($3^3$ neighbor voxels
are considered in the  $\Lstar$, see Paper I for details) with Ng acceleration
is virtually identical to that of  the 1D tri-diagonal $\Lstar$ case, cf.
Fig.~\ref{fig:cont:convergence}.  The importance of a non-local $\Lstar$ for
good convergence in the 3D case can be seen by comparing the results obtained
with a local (diagonal) $\Lstar$. Its convergence with Ng acceleration is
actually worse than that of the non-local $\Lstar$ without Ng acceleration.
The excellent convergence of the non-local $\Lstar$ can be used to reduce the
computing time by using a less stringent convergence criterion that still gives
results accurate enough for a particular application. One has to
be careful to be not too aggressive  in relaxing the convergence criterion. This
is highlighted in Fig.~\ref{fig:cont:relax} where the results for the large
test case are shown for different combinations of $\Lstar$ with Ng acceleration
and difference convergence limits. It is obvious that the combination of Ng
acceleration and a relatively large convergence limit with a local (diagonal)
$\Lstar$  is not acceptable, whereas the same convergence limit ($10^{-2}$
maximum relative change of $J$) with a non-local (nearest neighbor) $\Lstar$
gives usable results even without Ng acceleration. Setting a convergence limit
of $10^{-4}$ gives reasonable results  in all test cases and can reduce
computing time by about 50\%.

{With 128 MPI processes on the SGI ICE system of the
H\"ochstleistungs Rechenzentrum Nord (HLRN) the continuum test with
$\epsilon=10^{-4}$ and $n_r = n_\phi = 65$ and $n_\theta=33$ spatial points and
$256^2$ solid angle points requires 16 iterations and 1543\,s wallclock time,
where a formal solution takes about 80\,s, the construction of the non-local
$\Lstar$ about 140\,s and about 1.6\,s are used for the solution of the
operator splitting step. For the same case with $64^2$ solid angles the total
wallclock time is 230\,s. The memory requirements are about 45MB/process.}

\subsubsection{Line tests}

\begin{figure}
\centering
\resizebox{\hsize}{!}{\includegraphics[angle=90]{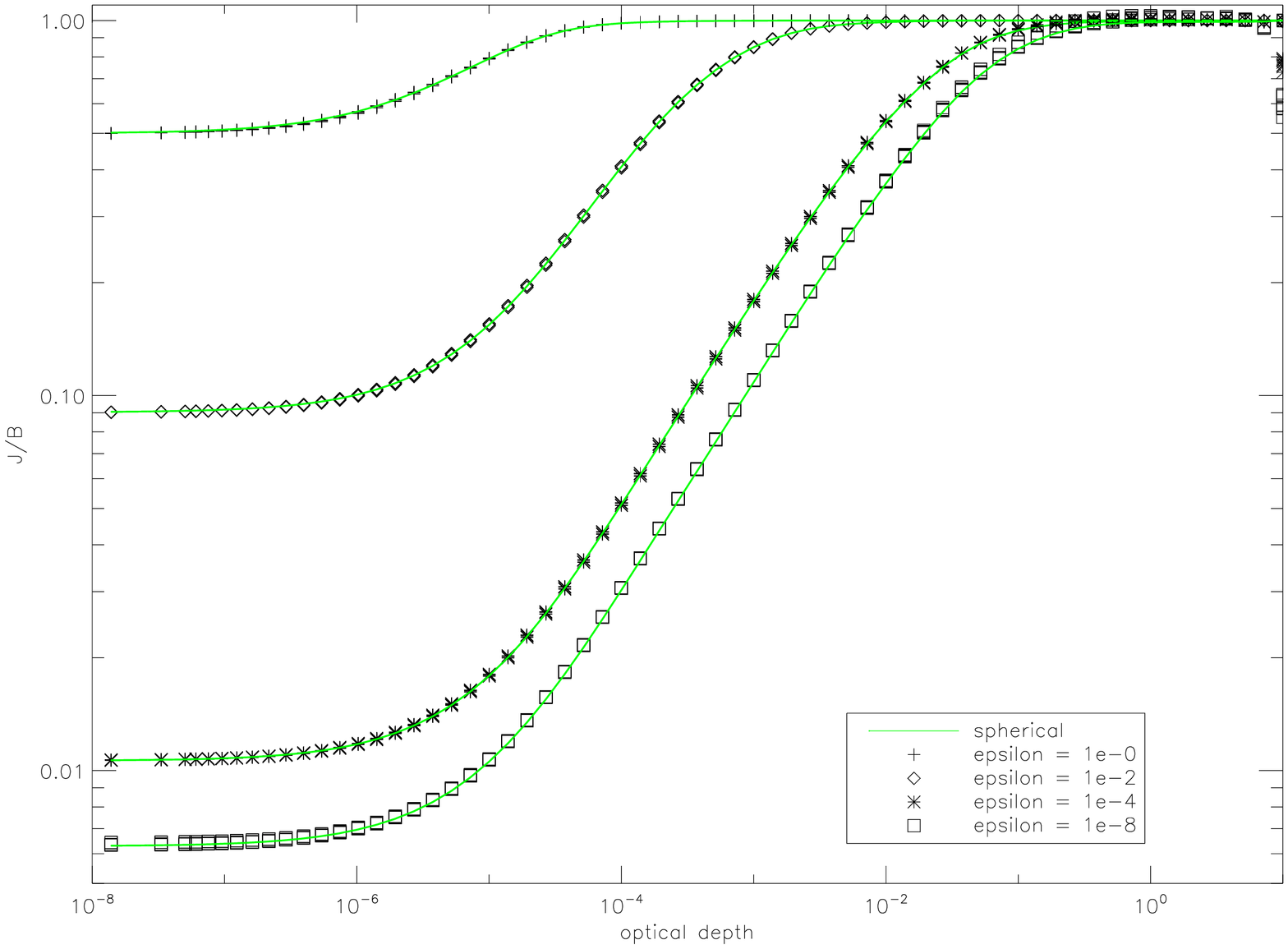}}
\caption{\label{fig:2lvl} The ratio of the line mean intensity $\Jbar$ and the
Planck function $B$ as function of the radial optical depth $\tstd$ for the
'large' test case in spherical coordinates for different values of $\epsilon_l$. The optical depth is
measured along the radius $r$. 
}
\end{figure}

In Fig.~\ref{fig:2lvl} we show the results of a comparison of the 
ratio  of the line mean intensities $\Jbar$ and the Planck function $B$
for the 'large' test case and various values of $\epsilon_l$. 
The 1D comparison models assume a diffusive inner boundary condition,
whereas the 3D models use $I=B$ as the boundary condition for the inner
boundary. Therefore, the models will differ close 
to the inner boundary. Away from the inner boundary the results 
of the 1D and 3D calculations agree very well with each other. The number
of iterations required to reach a prescribed accuracy is similar to the 
continuum tests described above. This shows that the quality of the 3D 
calculations match the 1D results, which is very important for future applications.

\begin{figure}
\centering
\resizebox{\hsize}{!}{\includegraphics[angle=00]{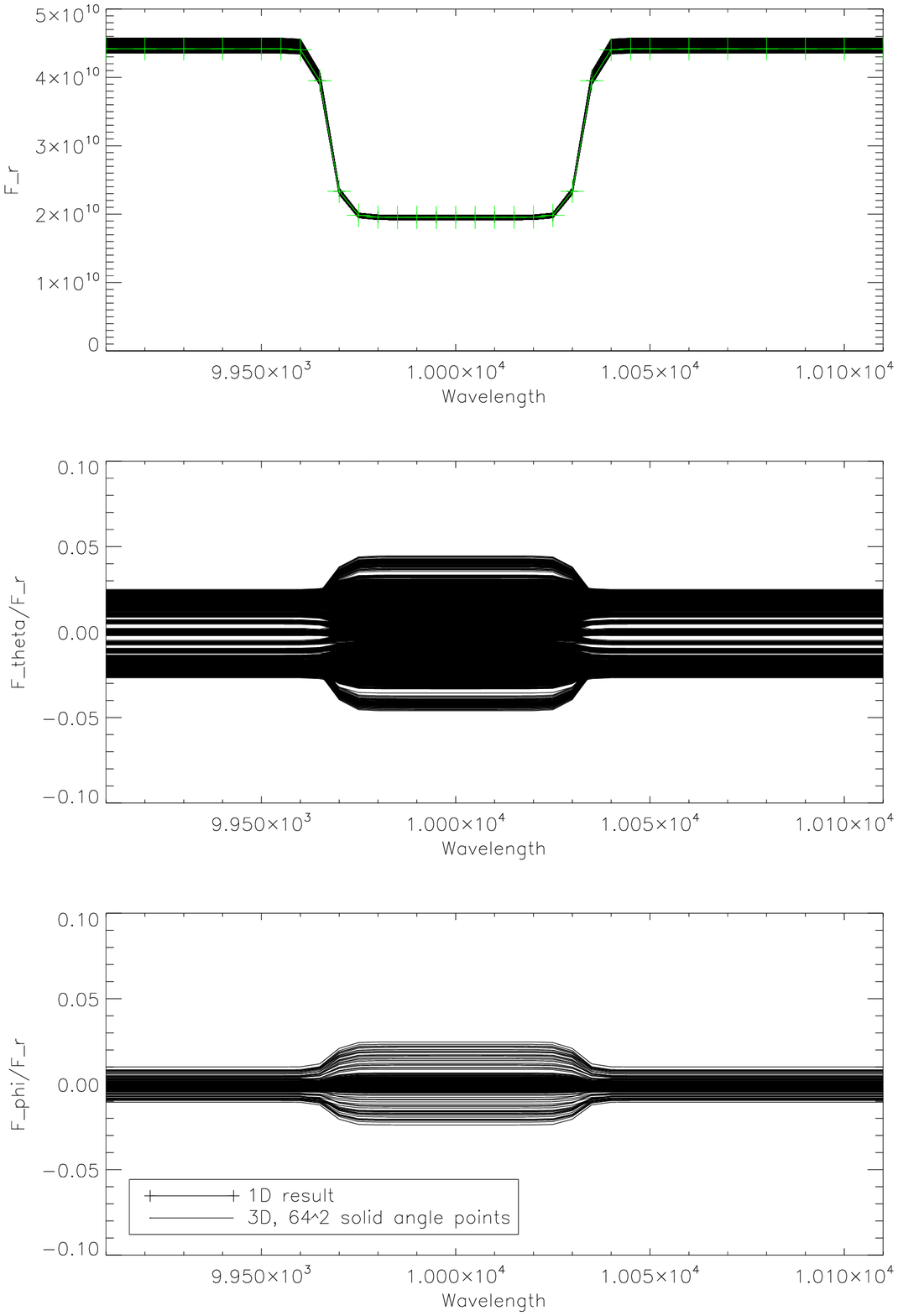}}
\caption{\label{fig:flux:eps0:om064} Flux vectors in the outermost voxels
for the line transfer test in spherical coordinates with $\epsilon_l=1$ and $64^2$ solid angle points.
The top panel shows the radial component $F_r$ of the voxel flux vector compared
to the results for the 1D comparison calculation. The middle and bottom panels 
show the ratio of the polar $F_\theta$ and azimuthal $F_\phi$ flux components to $F_r$, respectively.}
\end{figure}

\begin{figure}
\centering
\resizebox{\hsize}{!}{\includegraphics[angle=00]{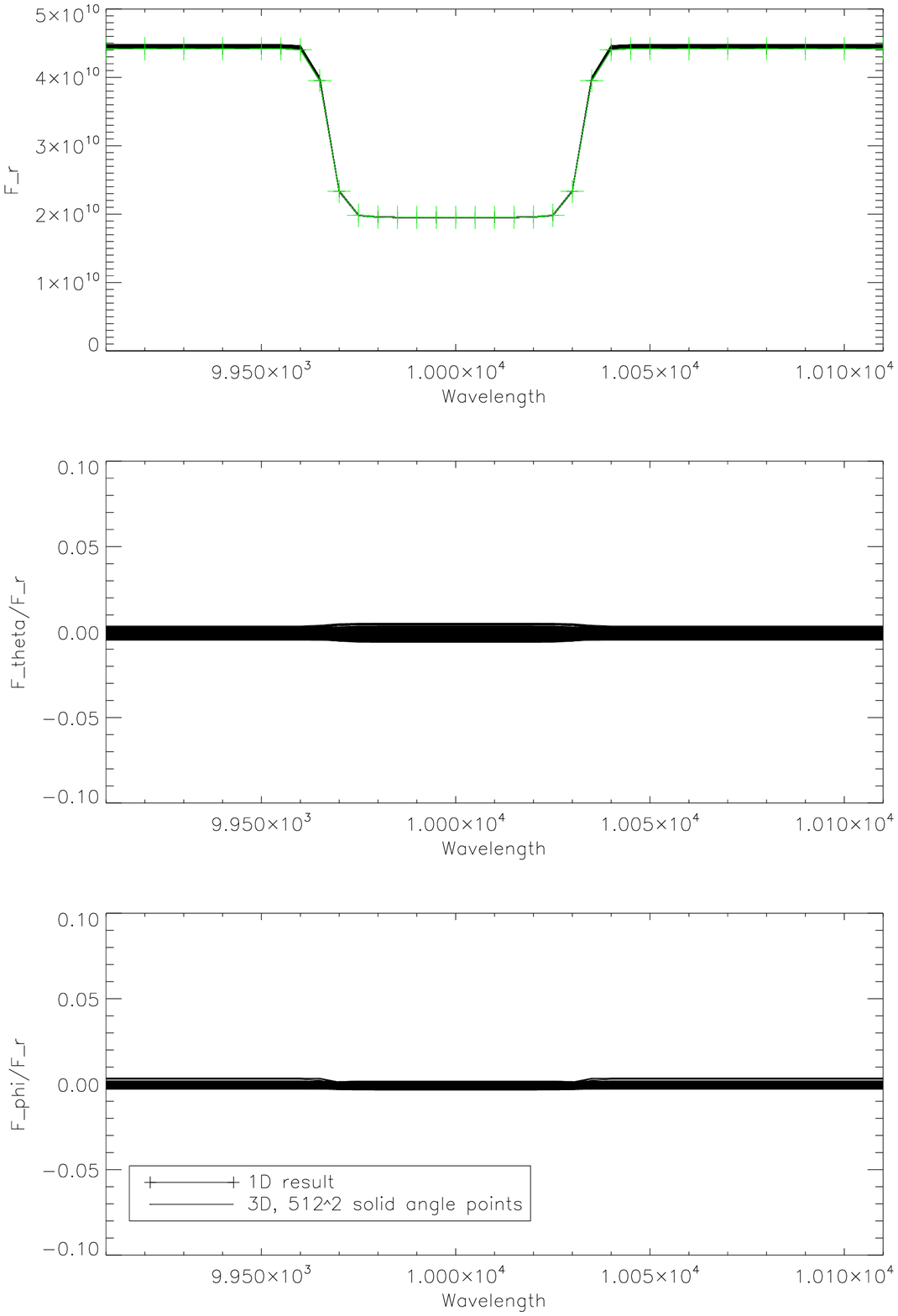}}
\caption{\label{fig:flux:eps0:om512} Flux vectors in the outermost voxels
for the line transfer test in spherical coordinates with $\epsilon_l=1$ and $512^2$ solid angle points.
The top panel shows the radial component $F_r$ of the voxel flux vector compared
to the results for the 1D comparison calculation. The middle and bottom panels 
show the ratio of the polar $F_\theta$ and azimuthal $F_\phi$ flux components to $F_r$, respectively.}
\end{figure}

\begin{figure}
\centering
\resizebox{\hsize}{!}{\includegraphics[angle=90]{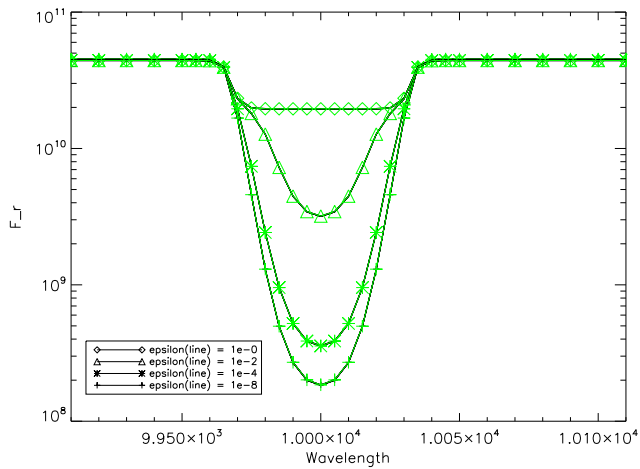}}
\caption{\label{fig:flux:eps:combined} Radial components $F_r$ of the flux vectors in the outermost voxels
for the line transfer tests in spherical coordinates with $\epsilon_l=1, 10^{-2}, 10^{-4}, 10^{-8}$ and $256^2$ solid angle points.
The results of the 1D calculations  are shown for comparison.}
\end{figure}

In Figs.~\ref{fig:flux:eps0:om064} and \ref{fig:flux:eps0:om512} we show the
flux vectors of the outermost voxels (outermost radius, all $(\theta_c,\phi_c)$
points) for the line transfer case with $\epsilon_l=1$ and for different
numbers of solid angle points. In the test case, the radial component of the
flux vector $F_r$ should be the only non-zero component as in spherical
symmetry both $F_\theta$ and $F_\phi$ are zero. The figures show that this is
indeed the case if a large enough number of solid angle points is used in the
calculation, if the number is too small, $F_\theta$ and $F_\phi$ can have
errors of about 10\% and $F_r$ scatters considerably around the 1D comparison.
For $512^2$ solid angle points, the scatter of $F_r$ is too small to be seen in
the plot and $F_\theta$ and $F_\phi$ are smaller than 1\% of $F_r$ over the
whole surface of the spherical grid.  This highlights the necessity of having
enough solid angle points to obtain good accuracy.
Fig.~\ref{fig:flux:eps:combined} shows $F_r$ compared to the 1D results for a
number of values of $\epsilon_l$. In all cases the results agree very well with
each other, showing the good numerical accuracy of the 3D code.
{The results of these tests are valid only for the test 
cases, therefore, a similar test should be performed for each ``real'' 3D
configuration individually in order to make sure that enough solid angles are
used for the accuracy requirements of each individual problem.}

{With 512 MPI processes on the SGI ICE system of the HLRN the line
test with $\epsilon_l=10^{-4}$ and $n_r = n_\phi = 65$ and $n_\theta=33$
spatial points, $32$ wavelength points and $256^2$ solid angle points requires
30 iterations and 20287\,s wallclock time and  for the same case with $64^2$
solid angles the total wallclock time is 1400\,s. In both cases we used
32 ``clusters'' with 16 MPI processes each so that for each wavelength 16
MPI processes work on a formal solution (see paper II for details).}

\section{Cylindrical coordinates}

The setup for the test of the code in cylindrical coordinates
is essentially identical to the tests of the spherical coordinate
system version. However, describing a spherically symmetric objects
in cylindrical coordinates is non-optimal and we expect results 
comparable to those presented in papers I and II. The cylindrical coordinate
system version of the code allows for different  maps in $\rho$ and $z$
for flexibility describing objects with nearly cylindrical 
symmetry.

The differences in the basic  setup compared to the spherical coordinates
tests are 
\begin{enumerate}
\item inner radius $r_c=10^{10}\,$cm, outer radius $\rout = 11\alog{10}\,$cm.
\item linear radial grid.
\item Continuum extinction $\chi_c = C/r^5$, with the constant $C$
fixed by the radius and optical depth grids, for the continuum tests,
$\chi_c = C$ for the line transfer tests.
\end{enumerate}
All other test parameters are the same as above.

\subsection{Results}

\begin{figure}
\centering
\resizebox{\hsize}{!}{\includegraphics[angle=90]{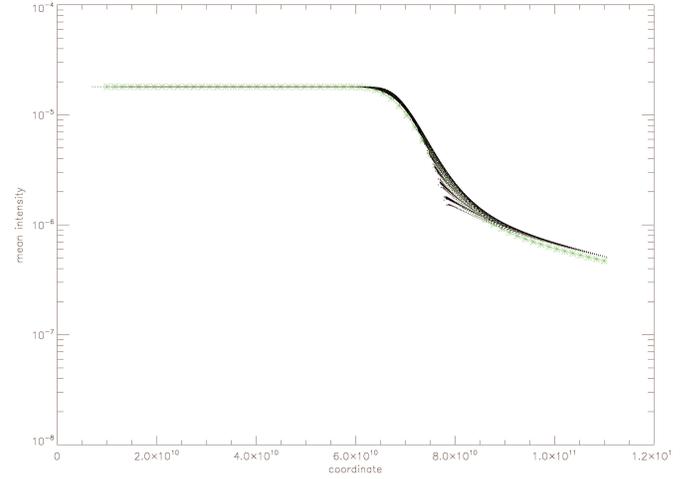}}
\caption{\label{fig:cyl} Results  for an example of a grey radiative transfer
calculation for a sphere in cylindrical 3D coordinates for $\epsilon=10^{-2}$
with $n_\rho=n_\theta=n_z=64$ and $64^2$ solid angle points.  The results of the
1D calculations  are shown for comparison with $*$ symbols.}
\end{figure}

For the cylindrical coordinate tests we show just the results
of a standard test discussed in Paper I and above for the 
spherical case with $\epsilon=10^{-2}$. The results for the continuum test case are presented
in Fig.~\ref{fig:cyl}. The results agree reasonably well with the 
1D test case, comparable to the results described in Paper I. The
reason for this is that describing a sphere in cylindrical
coordinates is not an ideal use of this coordinate
system, therefore much higher grid resolution is required 
to reach reasonable accuracy (the results for grids with 
 $n_\rho=n_\theta=n_z=64$ are substantially worse).
As for the test cases in spherical coordinates the resolution in
solid angles is very important for accurate results. The convergence 
behavior is the same as in the case of spherical coordinates.

\section{Conclusions}

We have discussed the extension of our 3D framework to 3D
spherical and cylindrical coordinate systems. These coordinates
are better suited than Cartesian coordinates for problems
that are approximately spherical (e.g., supernova envelopes) 
or cylindrical (e.g., protoplanetary disks). The convergence
properties and the accuracy of the solutions are comparable to
the 1D solution for the test cases discussed here. These additional
coordinate systems allow for more general use with higher internal
accuracy while using less memory than the purely Cartesian coordinate system.

The next step in the development of the 3D framework is to add the
treatment of velocity fields in the observer's frame and the 
co-moving frame. The latter will be important for applications
with relativistic velocity fields, e.g., supernovae, while 
the former is interesting for modeling the spectra of convective
flow models.

\begin{acknowledgements}
  This work was supported in part by by NASA grants NAG5-3505 and
  NAG5-12127,  NSF grants AST-0307323, and
  AST-0707704, and US DOE Grant DE-FG02-07ER41517, as well as 
  DFG GrK 1351 and SFB 676.
  Some of the
  calculations presented here were performed at the H\"ochstleistungs
  Rechenzentrum Nord (HLRN); at the NASA's Advanced Supercomputing
  Division's Project Columbia, at the Hamburger Sternwarte Apple G5
  and Delta Opteron clusters financially supported by the DFG and the
  State of Hamburg; and at the National Energy Research Supercomputer
  Center (NERSC), which is supported by the Office of Science of the
  U.S.  Department of Energy under Contract No. DE-AC03-76SF00098.  We
  thank all these institutions for a generous allocation of computer
  time.
\end{acknowledgements}

\bibliography{yeti,radtran,rte_paper2}

\clearpage

\end{document}